\documentclass[aip,longbibliography,floatfix]{revtex4-2}

\usepackage{amsmath,amssymb,bm}
\usepackage{graphicx}
\usepackage{xcolor}
\usepackage{hyperref}

\hypersetup{
  colorlinks=true,
  linkcolor=blue,
  citecolor=blue,
  urlcolor=blue
}

\newcommand{\dd}{\mathrm{d}}

\newcommand{\safeincludegraphics}[2][]{%
  \IfFileExists{#2}{\includegraphics[#1]{#2}}{%
    \fbox{\parbox{0.85\linewidth}{\centering Missing figure: \texttt{#2}}}%
  }%
}

\begin{document}

\title{Microphase Separation in Quorum-Sensing\\ Active Particles with Competing Sensing Scales}

\author{Michele Antonioli}
\affiliation{Department of Physics and Materials Science, University of Luxembourg, Esch-sur-Alzette L-1511, Luxembourg}

\author{Nicoletta Gnan}
\affiliation{ISC-CNR, Institute for Complex Systems, Piazzale A. Moro 2, 00185 Rome, Italy}
\email{nicoletta.gnan@cnr.it}

\author{Claudio Maggi}
\affiliation{NANOTEC-CNR, Soft and Living Matter Laboratory, Institute of
Nanotechnology, Piazzale A. Moro 5, 00185 Rome, Italy}
\email{claudio.maggi@cnr.it}

\date{\today}

\begin{abstract}

Standard quorum-sensing models in active matter exhibit collective phenomena such as motility-induced phase separation. Here, we show that incorporating competing sensing ranges---a minimal ingredient inspired by microbial communication---qualitatively changes this behavior, replacing macroscopic phase separation with self-organized microphases characterized by an emergent finite length scale.
Starting from the microscopic dynamics, we derive a coarse-grained field theory whose coefficients are explicitly related to the moments of the microscopic sensing function. This mapping enables a direct comparison between particle-based simulations and continuum theory, allowing the characteristic modulation and correlation lengths to be predicted directly from the microscopic interaction parameters. Two-dimensional numerical simulations confirm these predictions and reveal a transition from macrophase separation to finite-wavelength density modulations as the competition between sensing scales increases. For stronger competing interactions, the system develops a peculiar cluster phase with an interstitial percolating network, which is captured by a higher-order gradient expansion.
 Our results identify competing quorum-sensing interactions as a simple microscopic mechanism for generating tunable active microphases. 


\end{abstract}

\maketitle

\section{Introduction}
Self-organization is a fundamental characteristic of living systems, manifesting across all biological scales---from the spatial arrangement of cells within tissues to the large-scale organization of entire populations~\cite{koch1994biological}. In both multicellular organisms and microbial communities, coordinated behaviors are essential for regulating the spatial distribution of specialized cell types, facilitating tissue patterning~\cite{koromila2017broadly} and the formation of complex body plans~\cite{schauer2020zebrafish,sudderick2024periodic}. At the population level, self-organization is pivotal for achieving global order, enabling groups of cells or organisms to adapt to environmental changes, optimize the use of resources, and participate in collective behaviors that enhance survival.

In biological systems, self-organization can arise from different factors such as metabolic interactions~\cite{ackermann2023spatial}, competition for resources~\cite{wang2014trophic}, or motility-driven processes such as chemotaxis~\cite{colin2021multiple}. These mechanisms allow cells or organisms to respond dynamically to their surroundings, forming patterns and structures without external guidance~\cite{camazine2020self}. One of the most intriguing self-organization mechanisms in microbial populations is \emph{quorum sensing}, a communication process that allows cells to detect their population density and coordinate their behavior accordingly~\cite{waters2005quorum}. Quorum sensing is driven by the production, release, and detection of signaling molecules known as autoinducers. As the concentration of these molecules increases with population density, a threshold---referred to as a \emph{quorum}---is reached, triggering a collective response that can lead to the formation of biofilms~\cite{hammer2003quorum}, the expression of virulence factors~\cite{zhu2002quorum}, or the regulation of cell motility~\cite{daniels2004quorum}.
\newline In active matter physics, quorum sensing is frequently modeled as a direct coupling between particle motility and local density variations. This approach has led to the observation of collective aggregation in suspensions of light-activated colloids, where particle speed is locally controlled by density through feedback mechanisms~\cite{palacci2014light,bauerle2018self}. Analogously, aggregation has been observed in a quorum-sensing strategy where active particles adjust their velocities based on the direction of neighboring particles~\cite{lavergne2019group}. Theoretically, simple quorum-sensing models have proven effective in capturing complex collective behaviors, such as motility-induced phase separation (MIPS), which emerges from the effective attraction between particles resulting from their self-regulation of their motility~\cite{cates2015motility,solon2018generalized,gnan2022critical, zhou2024clustering}.  A major advance in the theoretical understanding of these systems was the
development of coarse-grained field theories showing that density-dependent motility naturally leads to effective Cahn--Hilliard descriptions and equilibrium-like phase
separation at large scales~\cite{stenhammar2013continuum,cates2015motility, maggi2022critical}. Moreover, the critical properties of quorum-sensing active particles have recently been characterized in detail, revealing equilibrium-like critical behavior associated with the onset of macrophase separation~\cite{gnan2022critical}.

Although active-matter models represent quorum sensing through simple
density-dependent motility rules, these descriptions are ultimately
inspired by communication mechanisms observed in microbial populations
~\cite{waters2005quorum,miller2001quorum}. In many bacterial species,
quorum sensing does not rely on a single signaling pathway. Instead,
cells integrate multiple chemical signals, allowing them to make
decisions based on both population density and environmental conditions
~\cite{ng2009bacterial,cornforth2014combinatorial}. Different signaling
molecules may possess distinct diffusion lengths and degradation rates, effectively introducing multiple sensing scales into the communication process. More generally, competing interactions acting at different length scales
are known to generate modulated phases and microphase separation in a wide
variety of equilibrium systems. A paradigmatic example is provided by
fluids with short-range attractive and long-range repulsive (SALR)
interactions~\cite{godfrin2014generalized}, where the competition between aggregation and long-range repulsion
suppresses complete phase separation and stabilizes finite-size clusters,
gels, and periodically modulated structures
~\cite{stradner2004equilibrium,toledano2009colloidal,zhuang2016recent}.
Similar finite-wavelength ordering phenomena arise in microemulsions,
where the competition between interfacial tension and amphiphilic
stabilization generates characteristic mesoscopic length scales
~\cite{Teubner1987}, as well as in block copolymers, where the
incompatibility between different polymer blocks competes with chain
connectivity and leads to ordered mesophases
~\cite{bates1990block}. These systems are commonly described by
Brazovskii-type field theories, in which the dominant instability occurs
at finite wavevector rather than at \(q=0\)
~\cite{brazovskii1975phase}. In addition, analogous microphase-forming
mechanisms have also been investigated in colloidal systems with
competing interactions through liquid-state theory, density-functional
approaches, and simulations
~\cite{imperio2006microphase,archer2008twodimensional,zhuang2016equilibrium}.

Recently, microphase separation has also attracted considerable attention in active models, where finite-size clusters and arrested phase separation have been shown to emerge from intrinsically non-equilibrium mechanisms that violate detailed balance, such as active interfacial currents and other irreversible contributions to the coarse-grained dynamics~\cite{Catesmicrophase,Joannymicrophase}. In these systems, the characteristic length scale is generated by non-equilibrium effects and does not generally admit an equilibrium free-energy description. This raises the question of whether active microphase separation can instead arise from mechanisms that admit a more direct equilibrium-like interpretation. Motivated by this question, we investigate a minimal extension of the standard quorum-sensing model in which particles perceive their local environment through two competing sensing ranges. While particles within a short-range neighborhood decrease their speed as the perceived density increases, particles located in an outer shell do the opposite. The standard quorum-sensing model is recovered as a limiting case when the outer contribution vanishes. Through a combination of numerical simulations, coarse-grained field theory, and renormalization analysis, we show that this simple modification qualitatively changes the collective behavior of the system. Rather than undergoing complete macrophase separation, the system develops microphase-separated states characterized by a finite emergent length scale. We further show that the coefficients of the effective field theory can be
expressed in terms of the microscopic parameters controlling the quorum-sensing interactions. This mapping establishes a direct bridge between particle-level simulations and mesoscopic theory, enabling quantitative predictions for the emergence of microphase separation and providing a rare example of an active-matter system in which microscopic and continuum descriptions can be systematically connected.

\section{Microscopic model}
\begin{figure}[t]
    \centering
    \safeincludegraphics[width=1.0\linewidth]{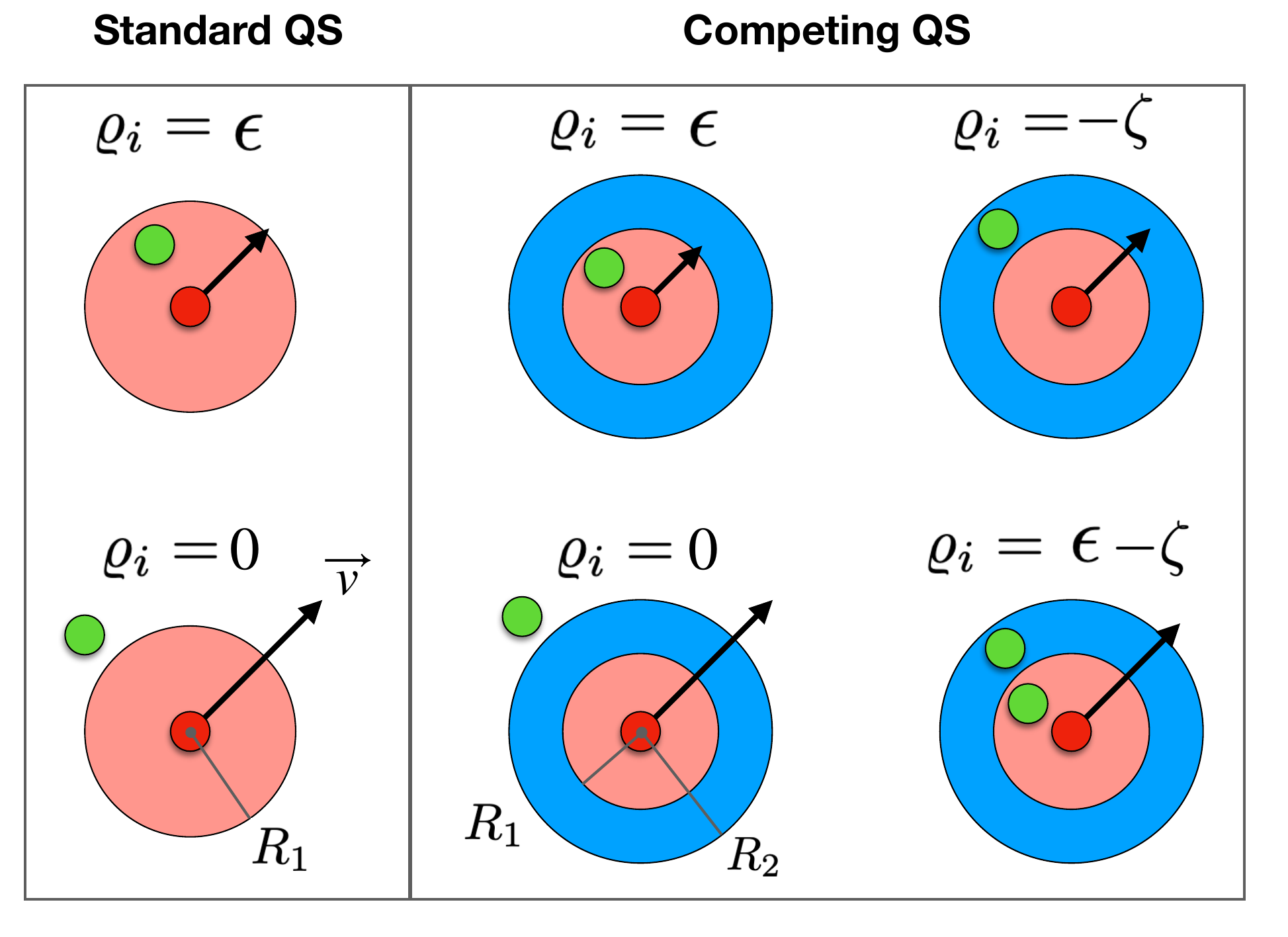}
    \caption{
    Schematic representation of the standard quorum sensing mechanism (left panel) compared to the model with competing interactions (right panel). The drawing shows how the sensed density $\varrho$ (entering the velocity kernel) is evaluated for a target particle (red) in the presence of neighboring particles (green). The black arrow indicates the velocity vector. In the left panel (standard quorum sensing), the red circle indicates the ``sensing'' neighborhood of radius $R_1$. When the red particle senses the neighboring particle, it slows down. In the right panel (model with competing interactions), the red circle indicates the inner neighborhood of radius $R_1$, while the blue disk indicates
    the outer neighborhood of radius $R_2$. A neighbor in the inner shell increases $\varrho_i$ and slows the red particle down, whereas a neighbor in the outer shell decreases $\varrho_i$ and speeds it up with respect to the inner-shell case.
    }
    \label{fig:QS_Scheme}
\end{figure}

We  consider a set $i=1,\ldots,N$ of Active Ornstein--Uhlenbeck particles (AOUP) whose equations of motion are
\begin{eqnarray}\label{eq:aoup}
\dot{\mathbf{r}}_{i}&=&s_i\boldsymbol{\eta}_{i},\nonumber\\
\tau \dot{\boldsymbol{\eta}}_i&=&-\boldsymbol{\eta}_i+\boldsymbol{\xi}_i .
\end{eqnarray}
Here $\mathbf{r}_i=(r_{i}^1,\ldots,r_{i}^d)$ is a $d$-dimensional position vector, $\boldsymbol{\eta}_i$ is the active force with persistence time $\tau$, and $\boldsymbol{\xi}_i$ is a delta-correlated noise satisfying
\begin{equation}
\langle \xi_i^{\alpha}(t)\xi_j^{\beta}(t')\rangle
=2D\delta_{ij}\delta_{\alpha\beta}\delta(t-t'),
\qquad
\langle\xi_i^{\alpha}\rangle=0 ,
\end{equation}
where $D$ is the active diffusion coefficient.
The scalar speed $s_i=s(\mathbf{r}_{i},\mathbf{r}_1,\ldots,\mathbf{r}_{N})$ allows particle $i$ to adjust its velocity according to the spatial configuration of the system. We model it as in~\cite{gnan2022critical}

\begin{equation}
 s(\mathbf{r}_{i},\mathbf{r}_1,\ldots,\mathbf{r}_{N}) = s(\varrho_i)=\frac{\Delta V}{1+\left(\varrho_i/\rho_0\right)^2}+V_0,
\label{eq:speed_kernel_main}
\end{equation}
where $\Delta V=V_{\textrm{max}}-V_0$ fixes the speed variation, $V_{\textrm{max}}$ is the free particle speed, $V_0$ is the residual speed at large perceived density, and $\rho_0$ sets the density scale. The key ingredient of the model is the perceived density
\begin{equation}
\varrho_i=\sum_{j\ne i}\gamma(|\mathbf{r}_i-\mathbf{r}_j|),
\label{eq:perceived_density_main}
\end{equation}
with a counting function
\begin{equation}
\gamma(r)=
\begin{cases}
\epsilon, & 0<r<R_1,\\
-\zeta, & R_1\le r<R_2,\\
0, & r\ge R_2 .
\end{cases}
\label{eq:counting_function}
\end{equation}
Particles inside an inner region $R_1$ contribute positively, with $\epsilon >0$, while particles in an outer shell between $R_1$ and $R_2$ contribute negatively, for $\zeta>0$. A sketch of the interaction mechanism is shown in Fig.~\ref{fig:QS_Scheme}.
Note that the standard quorum-sensing model is recovered as the limiting case
$\zeta=0$, for which the system undergoes ordinary macrophase separation.
The present model, therefore, allows us to continuously interpolate between
the previously studied MIPS regime and a new regime generated by competing sensing.

\section{Coarse-grained field theory}
Starting from the microscopic quorum-sensing dynamics, we derive an effective mesoscopic description for the coarse-grained density field $\rho(\mathbf r)$. For a single active particle in a spatially dependent speed field, the stationary probability density for the process of Eq.~\eqref{eq:aoup} is proportional to the inverse speed~\cite{caprini2022dynamics}:
\begin{equation}
\label{eq:stationary_probability}
p(\mathbf{r}) = \frac{1}{Z\,s(\mathbf{r},\mathbf{r}_1,\ldots,\mathbf{r}_{N})},
\end{equation}
where $Z=\int \dd\mathbf{r}\,[s(\mathbf{r},\mathbf{r}_1,\ldots,\mathbf{r}_{N})]^{-1}$.

By replacing the instantaneous density with the averaged one (mean-field approximation),
\begin{equation}
\rho(\mathbf{r})=N\langle \delta(\mathbf{r}-\mathbf{r}_i)\rangle=Np(\mathbf{r}),
\end{equation}
we obtain the following self-consistency relation~\cite{gnan2022critical}:
\begin{equation}
\ln \rho(\mathbf{r})+\ln s[\rho(\mathbf{r})]=\mu,
\label{eq:self_consistency_main}
\end{equation}
where
\begin{equation}
    \mu=\mu_{\rm I}+\mu_{\rm NI}
\end{equation}
plays the role of an effective chemical potential. The term $\mu_{\rm I}$ contains the contributions that can be written as the functional derivative of an effective Hamiltonian, while $\mu_{\rm NI}$ collects the non-integrable contributions. In the following, we neglect the non-integrable contributions and retain only the integrable part of the theory. We will show that this approximation already captures the finite-wavevector instability underlying microphase separation.

Expanding the density field around the homogeneous state of mean density $\rho_m$ (see Sec.~I of the supplementary material),
\begin{equation}
    \rho(\mathbf r)=\rho_m+\phi(\mathbf r),
\end{equation}
and retaining terms up to fourth order both in the field and in spatial gradients, we obtain the effective Hamiltonian~\footnote{In general, a cubic term should also be retained in the Landau expansion. Here, however, it is neglected since, as discussed below, it is expected to remain quantitatively small throughout the parameter regime explored in the numerical simulations.}
\begin{equation}
    \mathcal H[\phi]
    =
    \int \dd^d r
    \left[
    \frac{a_1}{2}\phi^2
    +
    \frac{b_2}{2}|\nabla \phi|^2
    +
    \frac{b_4}{2}(\nabla^2\phi)^2
    +
    \frac{a_3}{4}\phi^4
    \right].
    \label{eq:effective_hamiltonian}
\end{equation}
With competing sensing scales as in Eq.~(\ref{eq:counting_function}), the coefficient of the square-gradient term is not fixed in sign. In particular, when $b_2<0$, the homogeneous phase becomes unstable against density modulations. The higher-order stabilizing term $b_4(\nabla^2\phi)^2$, with $b_4>0$, suppresses short-wavelength fluctuations, thereby selecting a finite characteristic wavelength.

In Fourier space, upon normalizing the coefficient of the $q^4$ term, the effective action reads
\begin{eqnarray}
\label{eq:Hq}
    \mathcal H[\phi]
    &=&
\frac{1}{2}
\int_\mathbf{q}\, |\phi_\mathbf{q}|^{2}
\left(q^{4}+bq^{2}+c\right)\\
&+&
\frac{\lambda}{4!}
\int_{\mathbf{q}_1\ldots\mathbf{q}_4}
\phi_{\mathbf{q}_1}\phi_{\mathbf{q}_2}\phi_{\mathbf{q}_3}\phi_{\mathbf{q}_4}\,
(2\pi)^{d}\,\delta^{(d)}(\mathbf{q}_1+\mathbf{q}_2+\mathbf{q}_3+\mathbf{q}_4).
\nonumber
\end{eqnarray}
Here and in the following, $\int_{\mathbf{q}}\equiv\int \dd^{d}q/(2\pi)^{d}$ and $\int_{\mathbf{q}_1\ldots\mathbf{q}_4}$ denotes the corresponding integration over the four wavevectors. The quadratic term in Eq.~(\ref{eq:Hq}) yields the Gaussian propagator,
\begin{equation} \label{eq:prop}
\langle\phi_{\mathbf{q}}\phi^*_{\mathbf{q}'}\rangle=
\frac{(2\pi)^d\delta^{(d)}(\mathbf{q}-\mathbf{q}')}{q^{4}+bq^{2}+c},
\end{equation}
which, up to multiplicative constants,
corresponds to the structure factor of the theory ${S(q)\propto (q^4+b q^2+c)^{-1}}$, i.e., the well-known Teubner--Strey~\cite{Teubner1987} structure factor.

For $b<0$, the structure factor develops a maximum at a finite wavevector,
\begin{equation}
    q_* = \sqrt{-\frac{b}{2}},
\end{equation}
indicating enhanced density fluctuations at a preferred finite length scale.

\noindent Since ${S(q_*)\propto (c-b^2/4)^{-1}}$, the Gaussian theory predicts a divergence of the structure factor at
\begin{equation}
    c=\frac{b^2}{4},
\end{equation}
corresponding to the onset of the finite-wavevector instability associated with microphase separation.
It is interesting to relate this behavior to the generalized phase diagram of passive fluids with SALR interactions. In these systems, clustering was found to approximately follow the conditions under which the corresponding reference fluid, obtained by removing the long-range repulsion, undergoes liquid-vapor phase separation~\cite{godfrin2014generalized}.
A similar connection emerges in our model at the Gaussian level.
The bare mass \(c(\Delta V,\rho)\) is controlled by the zeroth moment $\gamma_0$ of the sensing function; hence, it coincides with that of the reference single-scale QS system~\cite{gnan2022critical}, whose spinodal is given by $c=0$.
The introduction of the outer sensing shell modifies the gradient coefficient $b$, which can become negative, shifting the instability threshold from $c=0$ to $c=b^2/4>0$. Therefore, the microphase-separation threshold remains connected to the demixing line of the reference system, although it occurs at a shifted value, similarly to passive SALR fluids.

\begin{figure}[t]
    \centering
    \safeincludegraphics[width=1.0\linewidth]{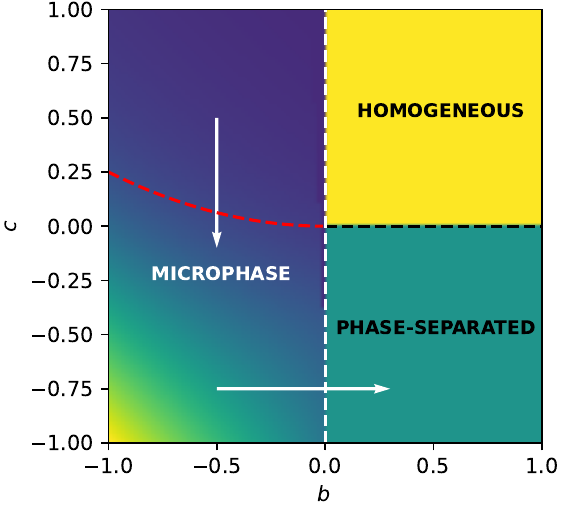}
    \caption{Mean-field phase diagram. The phase diagram is divided into four main regions: for $b>0$ and $c>0$, the system is homogeneous. Lowering $c$ and crossing the critical line at $c=0$ (black dashed line), the system phase separates. For $b<0$, the system forms a microphase, and the Gaussian theory predicts that, by lowering $c$ (vertical arrow), the system undergoes a phase transition at the critical line $c=b^2/4$ (red dashed line), where $\xi$ diverges. Differently, the renormalized theory always predicts a finite $\xi$, which smoothly increases upon lowering $c$ and $b$. The colormap represents the values of the renormalized $\xi$ from low (blue) to high (yellow).
    By increasing $b$ at fixed $c<0$ (horizontal arrow), the system undergoes a phase transition from a microphase to a full phase separation as the $b=0$ line is crossed (white dashed line). }
    \label{fig:PHD}
\end{figure}
Following Teubner and Strey~\cite{Teubner1987}, the spatial correlations can be characterized in terms of two emergent length scales,
\begin{equation}
\label{eq:xid}
    \xi
    =
    \frac{1}{
    \sqrt{
    \frac{\sqrt{c}}{2}+\frac{b}{4}
    }}, \; \; d
    =
    \frac{2\pi}{
    \sqrt{
    \frac{\sqrt{c}}{2}-\frac{b}{4}
    }},
\end{equation}
arising from an oscillatory decay of the spatial correlation function, which in three  dimensions is~\cite{levin1992renormalization}
\begin{equation}
    g(r-r')
    \sim
    \frac{e^{-|r-r'|/\xi}}{|r-r'|}
    \sin\!\left(
    \frac{2\pi |r-r'|}{d}
    \right).
\end{equation}
The quantity $\xi$ is the correlation length and controls the spatial extension of density fluctuations, while $d$ sets the characteristic periodicity of the modulated phase, namely, the typical spacing between neighboring high-density regions.

Within the Gaussian theory, the phase diagram is divided into four main regions (Fig.~\ref{fig:PHD}): for $b>0$, one recovers the standard picture where, for $c>0$, the system is homogeneous, while lowering $c$ and crossing the critical line at $c=0$, the system phase separates (this is the regime studied in~\cite{gnan2022critical}).
We note that the region $b<0$ and $c>b^2/4$, located above the red dashed line in Fig.~\ref{fig:PHD}, still belongs to the homogeneous phase. In this region, the Gaussian structure factor has a finite maximum at ${q^*=\sqrt{-b/2}}$,
indicating enhanced density fluctuations at the characteristic wavelength $2\pi/q^*$. However, these fluctuations remain spatially correlated only over a finite distance and do not produce persistent long-range modulated order. Upon decreasing $c$ at fixed $b<0$, the Gaussian theory predicts an instability at $c=b^2/4$, where the correlation length $\xi$ and the height of the structure-factor peak $S(q^*)$ diverge. However, it is known that, in three dimensions, the one-loop renormalized theory prevents the renormalized mass from reaching this threshold, and the correlation length, therefore, remains finite~\cite{levin1992renormalization}.
\begin{figure*}[t]
    \centering
 \safeincludegraphics[width=1.0\textwidth]{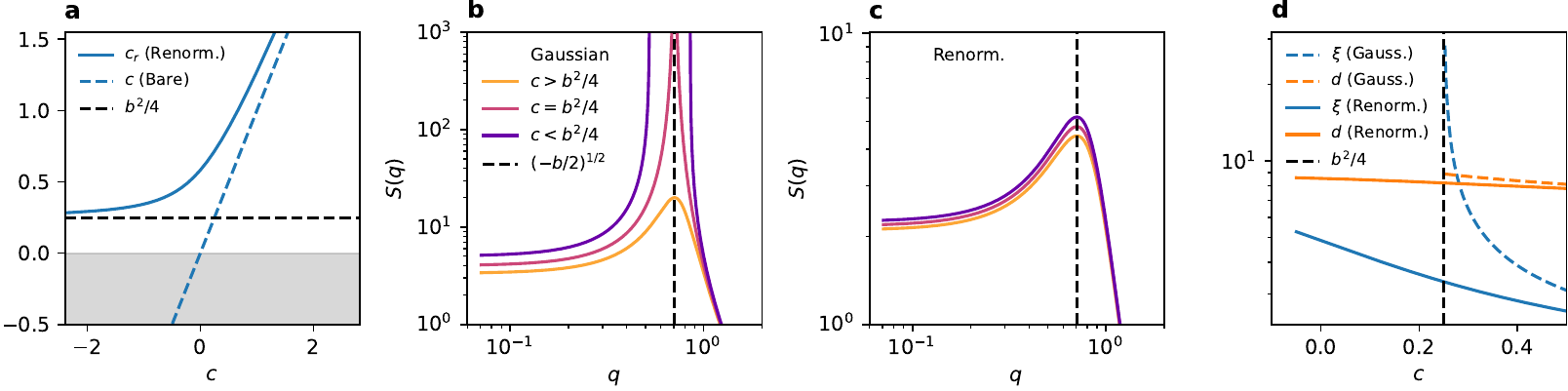}
 \vspace{-0.7cm}
    \caption{(\textbf{a}) Renormalized mass $c_r$ as a function of the bare mass $c$ (full line). $c_r$ is always above the critical value $b^2/4$ (dashed horizontal line) even for $c<0$ (gray area).
    (\textbf{b}) Gaussian structure factor at different values of the mass parameter $c$ (see legend). As the threshold value $c=b^2/4$ is approached, the peak at finite wavevector $q_*=\sqrt{-b/2}$ grows and eventually diverges.  (\textbf{c}) One-loop renormalized structure factor. In the renormalized theory, the bare mass $c$ is replaced by the self-consistent renormalized mass $c_r$, which regularizes the Gaussian divergence and keeps the structure factor finite even close to the instability threshold. (\textbf{d}) Characteristic length scales ($\xi$ and $d$) predicted by the field theory: Gaussian approximation (dashed lines) and renormalized theory (full lines). The vertical dashed lines indicate the threshold values at which the Gaussian structure factor develops a pole at finite wavevector.}
    \label{fig:Panel1Theory}
\end{figure*}

We have found that this result holds also in two dimensions, obtaining the following equation for the one-loop renormalized mass $c_r$:

\begin{equation}
\label{eq:c_r}
f(c_r) = c-c_r + \frac{\lambda \left[\pi - 2\arctan\!\left(\frac{b}{\sqrt{4c_r-b^2}}\right)\right]}
{8\pi\sqrt{4c_r-b^2 }}=0.
\end{equation}

\noindent Since for $\lambda>0$, $b<0$, and $c \in \mathbb{R}$, we have ${\lim_{c_r\rightarrow\infty} f(c_r)=-\infty}$ and ${\lim_{c_r\rightarrow(b^2/4)^+} f(c_r)=+\infty}$, Eq.~(\ref{eq:c_r}) always has a solution $c_r>b^2/4$, as guaranteed by the intermediate value theorem. The numerical solution of Eq.~(\ref{eq:c_r}) is plotted in Fig.~\ref{fig:Panel1Theory}(a), showing that $c_r$ lies above $b^2/4$ even when the bare mass $c$ is negative. This implies that, while the Gaussian $S(q)$ diverges upon decreasing $c$ [as shown in Fig.~\ref{fig:Panel1Theory}(b)], the renormalized structure factor is regularized for any value of $c$ [see Fig.~\ref{fig:Panel1Theory}(c)]. The regularization is also evident from the dependence of $\xi$ on $c$, as shown in Fig.~\ref{fig:Panel1Theory}(d). While $\xi$ diverges at $c=b^2/4$ in the Gaussian theory, the renormalized correlation length increases smoothly and remains finite as $c$ is decreased. 
 Differently, the bare modulation length $d$ does not diverge, and its renormalized value is only slightly lower than the bare one, as shown in Fig.~\ref{fig:Panel1Theory}(d).

\section{Numerical simulations of the macrophase-microphase transition}
\label{sec:simulations}

We numerically integrate the microscopic equations of motion in
Eq.~\eqref{eq:aoup} in two dimensions with periodic boundary conditions. The
present model is built as a generalization of the standard single-scale
quorum-sensing model studied in Ref.~\cite{gnan2022critical}, where
motility-induced phase separation was characterized. 
In the single-scale case, the sensing radius was fixed to
$R=1.3$~\cite{gnan2022critical}. Here we use, unless otherwise stated,
an inner sensing radius $R_1=R$ and an outer sensing range $R_2=2R_1$. Most
simulations are performed with $\rho_0=1$ and at density $\rho/\rho_0=2.82$. This value is
chosen because, in the standard single-scale quorum-sensing model, increasing
the activity at this density drives the system through the critical point and into the MIPS region~\cite{gnan2022critical}. It, therefore,
provides a natural reference state from which to investigate how the
competing outer sensing shell modifies the usual macrophase-separation
scenario. Moreover, in this region, we expect cubic field-theoretical terms [neglected in Eq.~(\ref{eq:effective_hamiltonian})] to be small, making this choice of parameters suitable for testing the theory.
The parameters $\epsilon$ and $\zeta$ control, respectively, the positive
inner-shell and the negative outer-shell contributions to the perceived
density in Eq.~\eqref{eq:counting_function}. In addition to these parameters,
the microscopic dynamics is controlled by the motility difference $\Delta V$,
the residual velocity $V_0=1$ [see Eq.~(\ref{eq:speed_kernel_main})], the persistence time $\tau=1$, and the active
noise amplitude $D=1$.
Most simulations are performed using systems of ${N=15\times10^3}$ particles,
although larger system sizes up to ${N=150\times10^3}$ have also been
considered in order to check finite-size effects.
Each configuration is equilibrated for $10^8$ integration steps and subsequently sampled for an additional $3\times10^8$ steps using a time step
$\dd t =10^{-4}$.

The link between the microscopic simulations and the field theory is provided
by the spatial moments of the counting function, introduced in
the supplementary material [Eqs.~(S10) and (S11)]. In two dimensions, they read
\begin{equation}
\gamma_0
=
\pi
\left[
\epsilon R_1^2
-
\zeta\left(R_2^2-R_1^2\right)
\right],
\end{equation}
and
\begin{equation}
\gamma_2
=
\frac{\pi}{8}
\left[
\epsilon R_1^4
-
\zeta\left(R_2^4-R_1^4\right)
\right].
\end{equation}
While $\gamma_0$ fixes the local contribution to the perceived density,
$\gamma_2$ controls the leading gradient correction and is, therefore, the
microscopic parameter associated with the square-gradient term of the
coarse-grained Hamiltonian, i.e., ${b\propto -\gamma_2/\gamma_4}$ in Eq.~(\ref{eq:Hq}). Therefore, according to the theory, by varying $\gamma_2$ we can change the effective surface tension.

To test this scenario, we fix $\gamma_0=1$ and tune $\gamma_2$ by changing the values of $\epsilon$ and $\zeta$ of the inner and outer sensing regions. All the remaining parameters ($\rho$ and $\Delta V$) are kept fixed in the simulations. The idea is to move continuously from the standard quorum-sensing limit, where the inner contribution dominates and the system undergoes macrophase separation, to the competing-sensing regime, where the
outer shell frustrates coarsening and selects a finite modulation length. Note that this corresponds to moving along the horizontal arrow in the schematic phase diagram of Fig.~\ref{fig:PHD}.
\begin{figure*}[t]
\centering
\safeincludegraphics[width=0.99\textwidth]{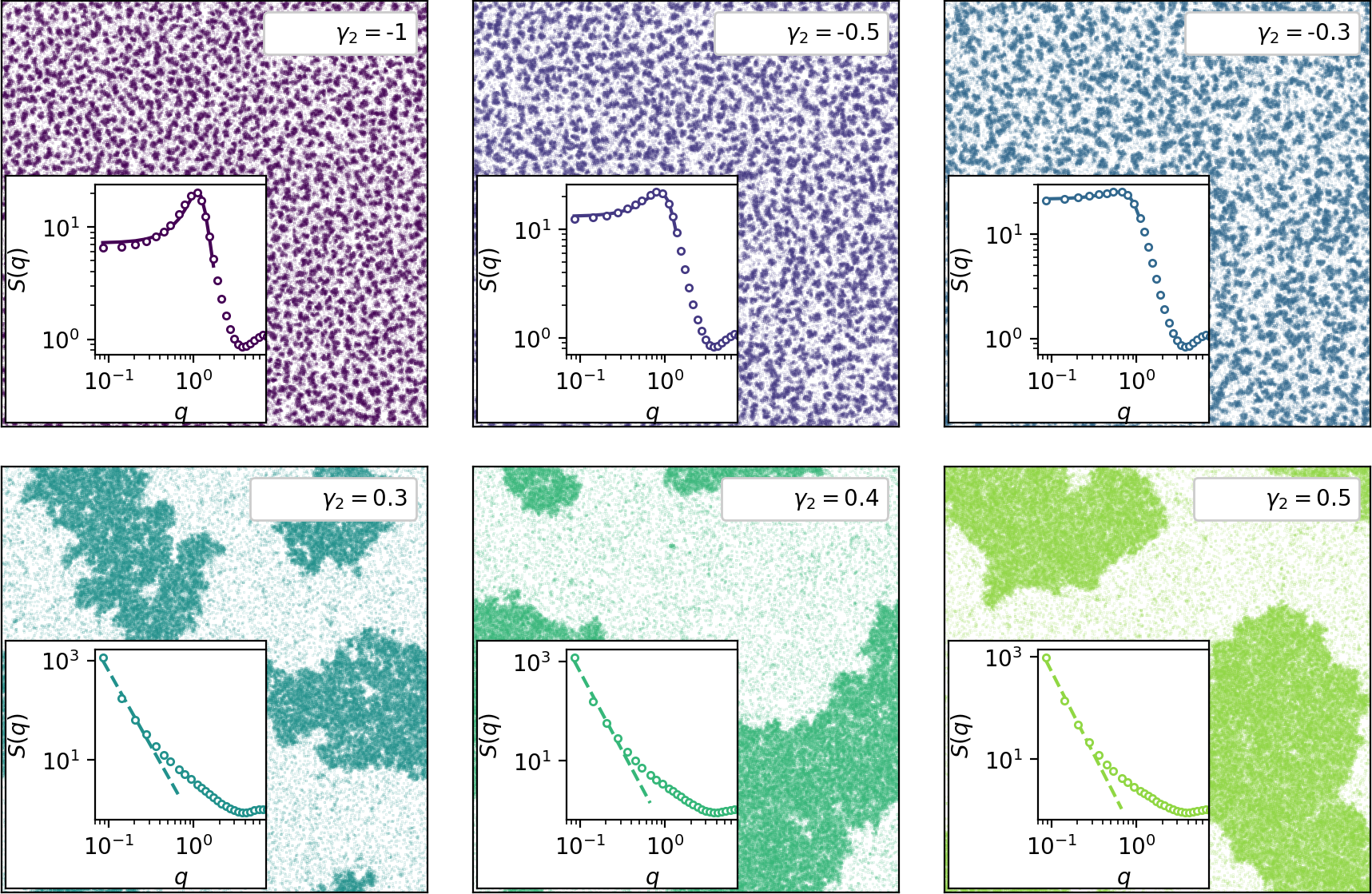}
\caption{
Representative steady-state configurations obtained from numerical
simulations of the competing quorum-sensing model at fixed parameters
$\rho/\rho_0=2.82$, $\Delta V/V_0=30$, and $\gamma_0=1$, for different values
of the parameter $\gamma_2$ (see legend).
The size of the system is $N=150\times10^3$ particles.
As $\gamma_2$ is increased, the system progressively crosses over from
microphase-separated states characterized by finite-size dense domains (top panels) to
macrophase-separated configurations (bottom panels) displaying large-scale coarsening.
For the parameters used here, varying $\gamma_2$ amounts to varying $\epsilon$ and $\zeta$ along the line $\zeta=(\gamma_0 R_1^2-8\gamma_2)/(12\pi R_1^4)$. The outer weight therefore vanishes at $\gamma_2=\gamma_0 R_1^2/8\simeq0.21$, and the three configurations with the largest $\gamma_2$ correspond to small negative $\zeta$, i.e., an outer shell that reinforces rather than opposes the inner one.
Insets: corresponding static structure factors $S(q)$.
For negative $\gamma_2$, the structure factor develops a pronounced peak at finite wavevector, signaling
the emergence of a characteristic modulation length associated with
microphase separation.
In this regime, the $S(q)$ is well fitted by ${S(q)\propto (q^4+b q^2+c)^{-1}}$ (continuous curve).
Increasing $\gamma_2$ suppresses the finite-$q$ peak
and shifts the dominant fluctuations toward small wavevectors, consistent
with the recovery of standard macrophase separation, where the $S(q)$ at low $q$ can be fitted by a power law (dashed line).
}
\label{fig:simulation_snapshots}
\end{figure*}
In Fig.~\ref{fig:simulation_snapshots} we show that this qualitative picture is clearly found in numerical simulations. These configurations show that the morphology of the system strongly depends on the value of $\gamma_2$: for
negative $\gamma_2$, the system does not coarsen into a single dense phase but instead develops finite-size clusters and modulated structures
characterized by a well-defined spacing. As $\gamma_2$ is increased
toward positive values,
the system crosses over to the large-scale coarsening
typical of standard macrophase separation.
This behavior is quantified by the static structure factor shown in the insets of Fig.~\ref{fig:simulation_snapshots}, computed as
\begin{equation}
S(q)=\frac{1}{N}
\left\langle
\left|
\sum_{j=1}^N
e^{-i\mathbf{q}\cdot\mathbf{r}_j}
\right|^2
\right\rangle,
\label{eq:numerical_structure_factor}
\end{equation}

\noindent which qualitatively changes upon varying $\gamma_2$. In the microphase-separated
regime, $S(q)$ develops a pronounced maximum at a finite wavevector. As shown in the top
panels of Fig.~\ref{fig:simulation_snapshots}, in this regime the numerical structure factor at small $q$
is well fitted by the following expression~\cite{Teubner1987}:

\begin{equation}
S(q)=\frac{A}{q^{4}+bq^{2}+c}.
\label{eq:sq}
\end{equation}
Upon
increasing $\gamma_2$, the finite-$q$ peak is progressively suppressed, and
the dominant fluctuations shift toward small wavevectors, consistent with
the recovery of macrophase separation.
The corresponding structure factors display a power-law regime, as shown in the insets of the bottom panels of Fig.~\ref{fig:simulation_snapshots}. For systems with sharp interfaces, Porod's law predicts~\cite{porod1951rontgenkleinwinkelstreuung,hansen2013theory}
\begin{equation}
S(q)\sim q^{-\alpha},
\qquad
\alpha=d+1,
\end{equation}
where $d$ is the spatial dimension. Therefore, in two dimensions, the expected exponent is $\alpha=3$. Fitting the numerical data gives
$\alpha=3.03\pm0.25$, $3.21\pm0.28$, and $3.26\pm0.30$ for
$\gamma_2=0.3$, $0.4$, and $0.5$, respectively, in good agreement with the Porod prediction.


Additional evidence on the nature of the macrophase--microphase
transition is reported in the supplementary material (Sec.~II\,B), where a finite-size
analysis of the Binder cumulant and the susceptibility suggests a weakly
first-order scenario associated with the competition between homogeneous
and finite-wavevector ordering.
Overall, the numerical results provide
direct evidence for the connection between the microscopic quorum-sensing
dynamics and the effective field-theoretical description. In particular, the observed crossover from macrophase to microphase separation confirms the
mapping between the microscopic parameters controlling the sensing function of Eq.~(\ref{eq:counting_function}) and the mesoscopic coefficients entering the coarse-grained Hamiltonian of Eq.~(\ref{eq:Hq}).

\section{Theory--simulation comparison in the microphase region}
\label{sec:theory_simulation}

\begin{figure*}[t]
\centering
\safeincludegraphics[width=0.999\textwidth]{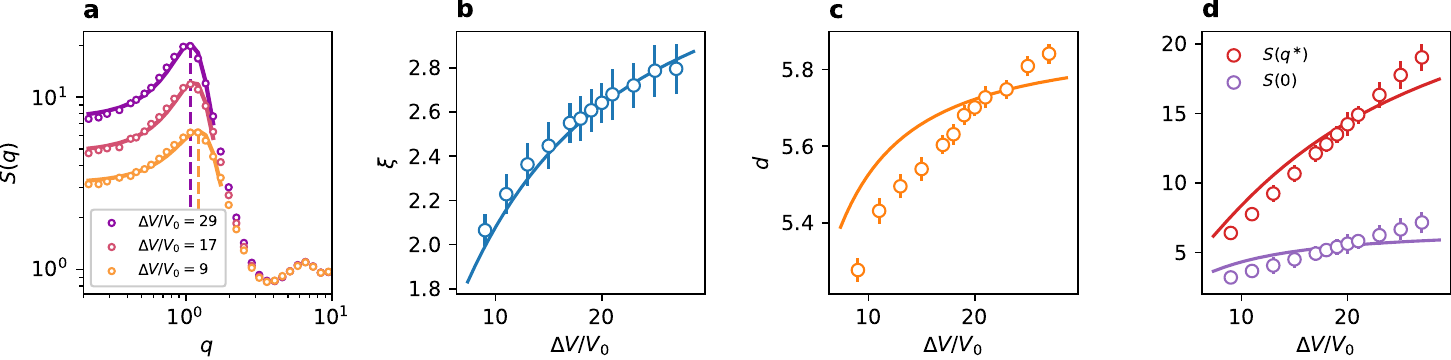}
\caption{
(\textbf{a}) Numerical structure factor in the microphase region at different values of the parameter $\Delta V$ (data points, see legend). Parameters are set to $N=15\times 10^3$ particles, $\rho/\rho_0=2.82$, and $\gamma_2/\gamma_0=-0.5$. Upon increasing $\Delta V$, $S(q)$ grows in amplitude, and its peak slightly shifts to lower $q$ (see vertical dashed lines). Around the peak, $S(q)$ can be fitted well with the Teubner--Strey form
$S(q)=A/(q^4+bq^2+c)$ (full lines) that allows us to extract the characteristic lengths $\xi$ and $d$. Panels (\textbf{b}) and (\textbf{c}) show, respectively, the correlation length $\xi$ and the modulation length $d$ as a function of $\Delta V$ (circles), where the error bars are estimated from the fits of the data in panel (a). The full lines in panels (b) and (c) represent a fit with the renormalized theory. Panel (\textbf{d}) shows the fitted values $S(q^{\ast})$ [the maximum of $S(q)$] and $S(0)$ as a function of $\Delta V$, while the full lines are the theoretical prediction adjusted by a single numerical prefactor.}
\label{fig:numerical_lengths}
\end{figure*}

To directly test the theory in the microphase regime, we run simulations at fixed sensing parameters ($\epsilon$ and $\zeta$) such that the system has $\gamma_2=\mathrm{const}<0$. We then fix $\rho=2.82\,\rho_0$ and vary the speed-change parameter $\Delta V$ between $9\,V_0$ and $29\,V_0$. With this choice, and using the mapping (see the supplementary material) from the microscopic parameters to the parameters of the theory [Eq.~(\ref{eq:Hq})], the gradient term $b$ is constant. Differently, the bare mass $c$, the renormalized mass $c_r$, and the quartic parameter $\lambda$ depend on the control parameters $\Delta V$ and $\rho$,

\begin{eqnarray}
\label{eq:mapp1}
b &=& \mathrm{const}<0,\\
c &=& c(\Delta V, \rho),\\
\lambda &=& \lambda(\Delta V, \rho),\\
c_r &=& c_r(\Delta V, \rho),
\label{eq:mapp4}
\end{eqnarray}

\noindent where $c_r$ depends on $\Delta V$ and $\rho$ through both $c$ and $\lambda$ [see Eq.~(\ref{eq:c_r})]. In particular, both $c$ and $c_r$ decrease monotonically with $\Delta V$. However, since $c_r$ always remains above the critical value $b^2/4$, the one-loop theory predicts that, as $c$ decreases with increasing $\Delta V$, the structure factor and the associated correlation length $\xi$ do not diverge (Fig.~\ref{fig:Panel1Theory}).
This prediction is confirmed by our numerical simulations, shown in Fig.~\ref{fig:numerical_lengths}(a), where the structure factors $S(q)$ grow in amplitude with increasing $\Delta V$ without diverging. The $S(q)$ data are well fitted at low $q$ by Eq.~(\ref{eq:sq}), from which we extract the characteristic lengths $\xi$ and $d$ [Eq.~(\ref{eq:xid})]. As shown in Figs.~\ref{fig:numerical_lengths}(b) and \ref{fig:numerical_lengths}(c), both $\xi$ and $d$ increase slowly with no sign of divergence, consistent with the renormalized theory.

 To test the theory more quantitatively, we exploit the fact that $\xi$ and $d$ depend on the parameters $b$ and $c$, which are themselves functions of $\Delta V$ and $\rho$; the characteristic lengths can, therefore, be written as $\xi=\xi(\Delta V,\rho)$ and $d=d(\Delta V,\rho)$. These functions follow directly from combining the microscopic mapping [Eqs.~(\ref{eq:mapp1})--(\ref{eq:mapp4})] with Eq.~(\ref{eq:xid}), which in principle allows $\xi$ and $d$ to be computed directly from the microscopic parameters. It is well known, however, that mean-field (or even one-loop) calculations often yield incorrect quantitative results, as in the case of the critical temperature of the Ising model~\cite{stanley1973introduction}. This is confirmed by our previous results~\cite{gnan2022critical}, where the theoretical estimates of the critical parameters $\Delta V$ and $\rho$ were found to differ significantly from the numerical values of the quorum-sensing system: the theory was qualitatively correct but underestimated the critical values, with $\Delta V_c^\mathrm{sim}\approx 2.4\,\Delta V_c^\mathrm{theo}$ and $\rho_c^\mathrm{sim}\approx 1.6\,\rho_c^\mathrm{theo}$. In that case, the numerical phase-separation region could be approximately rescaled onto the theoretical one by dividing the control parameters by two scale factors, $\overline{\Delta V}$ and $\overline{\rho}$, i.e., by setting $\Delta V\rightarrow \Delta V/\overline{\Delta V}$ and $\rho\rightarrow \rho/\overline{\rho}$.

Motivated by this, we fit the $\xi$ and $d$ values in Figs.~\ref{fig:numerical_lengths}(b) and \ref{fig:numerical_lengths}(c) with the following formulas:

\begin{eqnarray}
\xi&=&\xi(b,c_r(\Delta V/\overline{\Delta V},\rho/\overline{\rho})),
\label{eq:xifit} \\
d&=&d(b,c_r(\Delta V/\overline{\Delta V},\rho/\overline{\rho})),
\label{eq:dfit}
\end{eqnarray}

\noindent treating $\overline{\Delta V}$ and $\overline{\rho}$ as free fitting parameters. We stress that in this fit all other microscopic parameters of the theory (such as the sensing ranges and values) are held fixed, and only the two scale parameters are varied. Note also that the fit must consistently use the renormalized mass $c_r$ rather than the bare mass $c$, since the latter yields a diverging $\xi$, as discussed above. As shown by the solid lines in Figs.~\ref{fig:numerical_lengths}(b) and \ref{fig:numerical_lengths}(c), the fitted theory reproduces well the values of $\xi$, while it captures only qualitatively the slow growth of $d$ (which, however, varies much more moderately than $\xi$ over this range of $\Delta V$).
The scale factors obtained from the fit are ${\overline{\Delta V}=(1.8\pm0.1)\,V_0}$ and ${\overline{\rho}=(2.60\pm0.02)\,\rho_0}$, consistent in order of magnitude with the values obtained in~\cite{gnan2022critical}.

Finally, we note that the peak position $q_*$ decreases
from $q_*\approx1.2$ to $q_*\approx1.1$ as
$\Delta V/V_0$ grows approximately from 9 to 29 [Fig.~\ref{fig:numerical_lengths}(a)], a modest but genuine downward shift. Our mapping predicts $b$, and hence
$q_*=\sqrt{-b/2}$, to be independent of $\Delta V$ along this scan, since
it retains only the leading contribution to $b$ from the second moment of
the sensing function. The observed shift, therefore,
reflects a weak, sub-leading dependence of $b$ on $\Delta V$ not captured
at this loop-order~\cite{levin1992renormalization}. We also note [Fig.~\ref{fig:numerical_lengths}(d)] that the maximum of $S(q)$ [denoted by $S(q_*)$] grows by a factor $\sim3$, while $S(0)$ (defined in Ref.~\footnote{%
This limit has to be interpreted as taken after the thermodynamic limit, i.e.,
$S(0)=\lim_{q\rightarrow 0}\lim_{N,L^2\rightarrow\infty} S(q)$ (with $N/L^2=\mathrm{const}$ and $L$ being the linear size of the system).
${S(0)=\lim_{q\rightarrow 0}S(q)}$ grows only by a factor $\sim2$. This suggests that, as $\Delta V$ increases, the
system drifts modestly toward the macrophase-separation region of the phase
diagram (i.e., the $b=0$ line in Fig.~\ref{fig:PHD}).
Over the range of $\Delta V$ explored here, however, $q_*$ remains well
away from zero and the peak amplitude continues to dominate over $S(0)$,
indicating that the system remains within the
microphase regime.}) grows only by a factor $\sim2$.


\section{Exotic morphologies and higher-order gradient expansion}
\label{sec:exotic_morphologies}

\begin{figure*}[t]
\centering
\safeincludegraphics[width=0.99\textwidth]{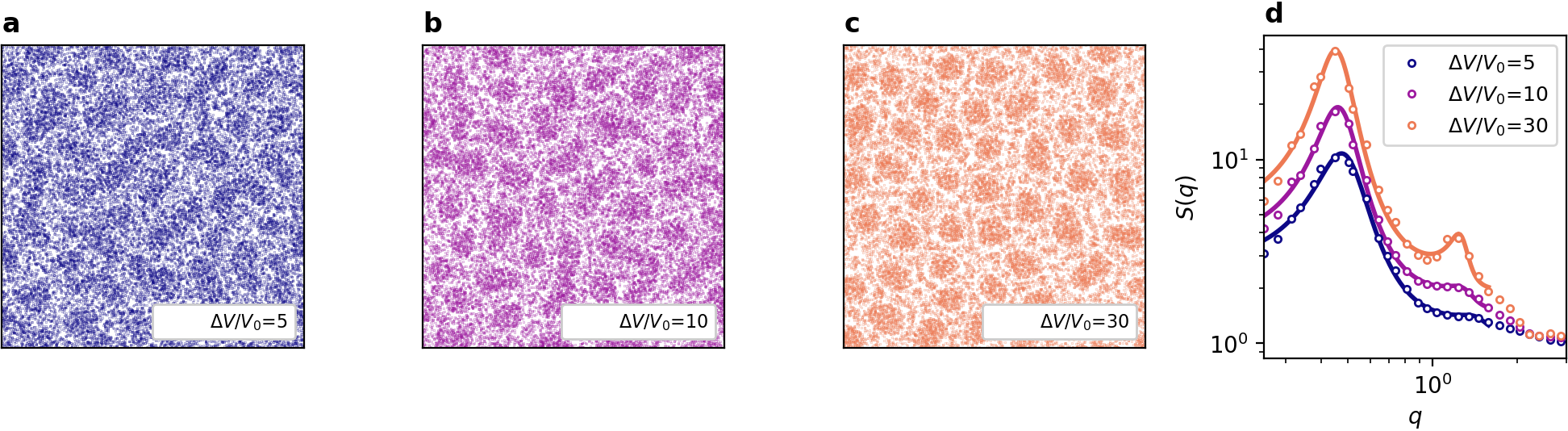}
\caption{(\textbf{a})--(\textbf{c}) Representative steady-state configurations obtained for larger sensing ranges and strongly negative $\gamma_2$. The system develops a complex morphology consisting of compact, dense clusters embedded within a dilute, percolating network. The simulation parameters are $N=30\times10^3$, $\rho/\rho_0=2.82$, and $R_2=2R_1$, with $R_1=4.0$, $\gamma_0=1$, and $\gamma_2/\gamma_0=-49.80$, while $\Delta V/V_0$ is varied as indicated in the panels.
(\textbf{d}) Corresponding static structure factors $S(q)$ (open circles). As $\Delta V/V_0$ increases, $S(q)$ progressively develops a secondary peak at larger $q$. The solid curves are fits obtained by extending the gradient expansion to include terms up to order $q^8$.}
\label{fig:foam}
\end{figure*}

By substantially changing the sensing ranges and values, we have found that the system's structure becomes richer than the simple cluster phase discussed above. In particular, for $R_1=4$, $R_2=8$, $\gamma_0=1$, and $\gamma_2=-49.8$, simulations reveal the formation of a fluid of clusters surrounded by a dilute percolating phase. This is shown in Figs.~\ref{fig:foam}(a)--\ref{fig:foam}(c). The emergence of this complex structure is accompanied by the emergence of a secondary peak in the $S(q)$ [see Fig.~\ref{fig:foam}(d)].
This observation further demonstrates that the cluster size is not directly determined by the microscopic interaction length, but instead emerges collectively through the competition between the two quorum-sensing interactions. These states are characterized by small-scale density modulations, exposing the limits of the low-$q$ field-theoretical expansion [see Eqs.~(\ref{eq:effective_hamiltonian}) and (\ref{eq:Hq})].
The mapping developed so far shows that the coefficients of the gradient expansion are directly related to the successive moments of the microscopic interaction function. While the first non-zero moments $\gamma_0$, $\gamma_2$, and $\gamma_4$ (see the supplementary material) are sufficient to predict the existence of a preferred modulation length, they cannot fully describe the structure of the modulated phase once additional features emerge at larger wavevectors. This interpretation is supported by the numerical structure factors of Fig.~\ref{fig:foam}(d). While these $S(q)$ could be fitted by the lowest-order theory of Eq.~(\ref{eq:sq}) only at small $q$, a larger range (including the secondary peak) can be fitted well with
\begin{equation}
S(q)=
\frac{A}{q^8 + h\,q^6+f\,q^4+ b\,q^2+c},
\end{equation}
where the coefficients have alternating signs (i.e., ${c>0}$, ${b<0}$, ${f>0}$, and ${h<0}$), which is the natural continuation of the gradient expansion of Eq.~({\ref{eq:Hq}}). The need to include the $q^6$ and $q^8$ terms, with the expected signs, indicates that higher moments of the microscopic sensing function must be retained in the effective free energy to describe the structure arising in this regime.

In the context of passive systems, we are not aware of a direct SALR analog of the morphology shown in Fig.~\ref{fig:foam}, where dense clusters are embedded within an interstitial dilute percolating network. Related, although structurally different, cluster-percolated states have been reported in passive SALR systems~\cite{tan2024quasi}. An important difference is that conventional SALR particles possess a hard or steep repulsive core, whereas our minimal quorum-sensing model does not include excluded-volume interactions. The resulting possibility of reaching very large local densities may favor the peculiar coexistence of compact clusters and an interstitial network observed here. Whether this morphology is specific to quorum-sensing systems remains an open question.

\noindent More generally, the competition between interactions acting over different length scales is known to generate a wide variety of  modulated phases in soft condensed matter in equilibrium~\cite
{barkan2014controlled, zhuang2016recent,zhuang2016equilibrium, caprini2018cluster,munao2022competition}. Consistent with this general picture, competing quorum-sensing interactions can generate a similarly rich phenomenology, including structures characterized by more than one modulation scale.

\section{Conclusions}
\label{sec:conclusions}
We have introduced a minimal extension of quorum-sensing active particles in which the local motility is regulated by two competing sensing ranges. Despite the simplicity of the microscopic interaction rule, the competition between the inner and outer sensing shells qualitatively changes the collective behavior of the system, suppressing ordinary macrophase separation and promoting the emergence of microphase-separated states characterized by a finite modulation length.

Starting directly from the microscopic dynamics, we derived a coarse-grained field theory whose coefficients are explicitly determined by the spatial moments of the quorum-sensing function. This microscopic-to-mesoscopic mapping provides a quantitative framework connecting particle-based simulations with continuum theory, allowing both the characteristic modulation length and the correlation length to be predicted directly from the microscopic interaction parameters. Such an explicit correspondence between microscopic dynamics and effective field theory remains relatively uncommon in active matter and provides a route toward quantitatively predictive continuum descriptions.

The resulting theory successfully captures the onset of finite-wavevector density modulations, and quantitatively reproduces the low-$q$ behavior of the numerical structure factors. The correlation and modulation lengths extracted from simulations are in substantial agreement with the theoretical predictions over a broad range of parameters, demonstrating that the emergence of active microphases can be understood directly from the microscopic sensing function.

We have found that, for stronger competing interactions, the phenomenology becomes richer, displaying a cluster phase surrounded by a percolating network.
These structures show small- and large-scale density modulations and the numerical structure factors further reveal that higher moments of the interaction function become progressively more important. This indicates that extending the gradient expansion beyond the leading-order approximation is necessary to quantitatively describe the full richness of the observed morphologies.

More broadly, our results identify competing quorum-sensing interactions as a simple microscopic mechanism capable of generating a wide variety of active microphases without introducing explicit competing pairwise interactions. Rather than being controlled by an effective interaction potential, the collective behavior is encoded in the spatial structure of the sensing function itself. This perspective naturally suggests that the hierarchy of spatial moments of the quorum-sensing function constitutes a set of physically meaningful control parameters for designing and classifying active microphases. We hope that this framework will stimulate further theoretical and experimental investigations of multiscale communication mechanisms in active matter and microbial systems and, more generally, contribute to the development of predictive continuum theories directly rooted in microscopic interaction rules.

\section*{SUPPLEMENTARY MATERIAL}

See the supplementary material for the detailed derivation of the coarse-grained effective Hamiltonian from the microscopic quorum-sensing dynamics, including the mapping between the moments of the sensing function and the field-theoretical coefficients. Additional numerical results include a state diagram as a function of density and motility contrast, representative configurations and structure factors, and a finite-size analysis of the macrophase--microphase transition based on the Binder cumulant and susceptibility, supporting a weakly first-order scenario.

\begin{acknowledgments}
N.G. and C.M. acknowledge the funding from the Italian Ministero dell'Universit\`{a}
e della Ricerca (MUR) under the program PRIN 2022, funded by the European Union---NextGenerationEU. Project Title: \emph{Motility and Communication in Active Matter} (MOCA)---Grant No. CUP B53D23004200006---Grant Assignment Decree No. 104 adopted on 02-02-2022.
\end{acknowledgments}
\bibliographystyle{apsrev4-2}
\bibliography{ArxivBibliography}

@book{stanley1973introduction,
  title     = {Introduction to Phase Transitions and Critical Phenomena},
  author    = {Stanley, H. Eugene},
  year      = {1971},
  publisher = {Oxford University Press},
  address   = {New York}
}

@article{porod1951rontgenkleinwinkelstreuung,
  title={Die R{\"o}ntgenkleinwinkelstreuung von dichtgepackten kolloiden systemen: I. Teil},
  author={Porod, G},
  journal={Kolloid-Zeitschrift},
  volume={124},
  number={2},
  pages={83--114},
  year={1951},
  publisher={Springer}
}

@book{hansen2013theory,
  title={Theory of simple liquids: with applications to soft matter},
  author={Hansen, Jean-Pierre and McDonald, Ian Ranald},
  year={2013},
  publisher={Academic press}
}

@article{levin1992renormalization,
  title={Renormalization of a Landau-Ginzburg-Wilson theory of microemulsion},
  author={Levin, Y and Mundy, CJ and Dawson, KA},
  journal={Physical Review A},
  volume={45},
  number={10},
  pages={7309},
  year={1992},
  publisher={APS}
}

@article{koch1994biological,
  title={Biological pattern formation: from basic mechanisms to complex structures},
  author={Koch, Andr{\'e}-Joseph and Meinhardt, Hans},
  journal={Reviews of modern physics},
  volume={66},
  number={4},
  pages={1481},
  year={1994},
  publisher={APS}
}

@article{koromila2017broadly,
  title={Broadly expressed repressors integrate patterning across orthogonal axes in embryos},
  author={Koromila, Theodora and Stathopoulos, Angelike},
  journal={Proceedings of the National Academy of Sciences},
  volume={114},
  number={31},
  pages={8295--8300},
  year={2017},
  publisher={National Acad Sciences}
}

@article{schauer2020zebrafish,
  title={Zebrafish embryonic explants undergo genetically encoded self-assembly},
  author={Schauer, Alexandra and Pinheiro, Diana and Hauschild, Robert and Heisenberg, Carl-Philipp},
  journal={Elife},
  volume={9},
  pages={e55190},
  year={2020},
  publisher={eLife Sciences Publications, Ltd}
}

@article{sudderick2024periodic,
  title={Periodic pattern formation during embryonic development},
  author={Sudderick, Zoe R and Glover, James D},
  journal={Biochemical Society Transactions},
  volume={52},
  number={1},
  pages={75--88},
  year={2024},
  publisher={Portland Press Ltd.}
}

@article{waters2005quorum,
  title={Quorum sensing: cell-to-cell communication in bacteria},
  author={Waters, Christopher M and Bassler, Bonnie L},
  journal={Annu. Rev. Cell Dev. Biol.},
  volume={21},
  number={1},
  pages={319--346},
  year={2005},
  publisher={Annual Reviews}
}

@article{miller2001quorum,
  author = {Miller, Melissa B. and Bassler, Bonnie L.},
  title = {Quorum sensing in bacteria},
  journal = {Annual Review of Microbiology},
  volume = {55},
  pages = {165--199},
  year = {2001}
}

@article{hammer2003quorum,
  title={Quorum sensing controls biofilm formation in Vibrio cholerae},
  author={Hammer, Brian K and Bassler, Bonnie L},
  journal={Molecular microbiology},
  volume={50},
  number={1},
  pages={101--104},
  year={2003},
  publisher={Wiley Online Library}
}

@article{zhu2002quorum,
  title={Quorum-sensing regulators control virulence gene expression in Vibrio cholerae},
  author={Zhu, Jun and Miller, Melissa B and Vance, Russell E and Dziejman, Michelle and Bassler, Bonnie L and Mekalanos, John J},
  journal={Proceedings of the National Academy of Sciences},
  volume={99},
  number={5},
  pages={3129--3134},
  year={2002},
  publisher={National Acad Sciences}
}

@article{daniels2004quorum,
  title={Quorum sensing and swarming migration in bacteria},
  author={Daniels, Ruth and Vanderleyden, Jos and Michiels, Jan},
  journal={FEMS microbiology reviews},
  volume={28},
  number={3},
  pages={261--289},
  year={2004},
  publisher={Blackwell Publishing Ltd Oxford, UK}
}

@article{cates2015motility,
  title={Motility-induced phase separation},
  author={Cates, Michael E and Tailleur, Julien},
  journal={Annu. Rev. Condens. Matter Phys.},
  volume={6},
  number={1},
  pages={219--244},
  year={2015},
  publisher={Annual Reviews}
}

@article{solon2018generalized,
  title={Generalized thermodynamics of phase equilibria in scalar active matter},
  author={Solon, Alexandre P and Stenhammar, Joakim and Cates, Michael E and Kafri, Yariv and Tailleur, Julien},
  journal={Physical Review E},
  volume={97},
  number={2},
  pages={020602},
  year={2018},
  publisher={APS}
}

@article{cornforth2014combinatorial,
  title={Combinatorial quorum sensing allows bacteria to resolve their social and physical environment},
  author={Cornforth, Daniel M and Popat, Roman and McNally, Luke and Gurney, James and Scott-Phillips, Thomas C and Ivens, Alasdair and Diggle, Stephen P and Brown, Sam P},
  journal={Proceedings of the National Academy of Sciences},
  volume={111},
  number={11},
  pages={4280--4284},
  year={2014},
  publisher={National Acad Sciences}
}

@article{bauerle2018self,
  title={Self-organization of active particles by quorum sensing rules},
  author={B{\"a}uerle, Tobias and Fischer, Andreas and Speck, Thomas and Bechinger, Clemens},
  journal={Nature communications},
  volume={9},
  number={1},
  pages={3232},
  year={2018},
  publisher={Nature Publishing Group UK London}
}

@article{palacci2014light,
  title={Light-activated self-propelled colloids},
  author={Palacci, J{\'e}r{\'e}mie and Sacanna, Stefano and Kim, S-H and Yi, G-R and Pine, David J and Chaikin, Paul M},
  journal={Philosophical Transactions of the Royal Society A: Mathematical, Physical and Engineering Sciences},
  volume={372},
  number={2029},
  pages={20130372},
  year={2014},
  publisher={The Royal Society Publishing}
}

@article{lavergne2019group,
  title={Group formation and cohesion of active particles with visual perception--dependent motility},
  author={Lavergne, Fran{\c{c}}ois A and Wendehenne, Hugo and B{\"a}uerle, Tobias and Bechinger, Clemens},
  journal={Science},
  volume={364},
  number={6435},
  pages={70--74},
  year={2019},
  publisher={American Association for the Advancement of Science}
}

@article{gnan2022critical,
  title={Critical behavior of quorum-sensing active particles},
  author={Gnan, Nicoletta and Maggi, Claudio},
  journal={Soft Matter},
  volume={18},
  number={39},
  pages={7654--7661},
  year={2022},
  publisher={Royal Society of Chemistry}
}

@book{camazine2020self,
  title     = {Self-Organization in Biological Systems},
  author    = {Camazine, Scott and Deneubourg, Jean-Louis and Franks, Nigel R.
               and Sneyd, James and Theraulaz, Guy and Bonabeau, Eric},
  year      = {2001},
  publisher = {Princeton University Press},
  address   = {Princeton, NJ}
}

@article{ackermann2023spatial,
  title={Spatial self-organization of metabolism in microbial systems: a matter of enzymes and chemicals},
  author={Ackermann, Martin and van Vliet, Simon and others},
  journal={Cell systems},
  volume={14},
  number={2},
  pages={98--108},
  year={2023},
  publisher={Elsevier}
}

@article{colin2021multiple,
  title={Multiple functions of flagellar motility and chemotaxis in bacterial physiology},
  author={Colin, Remy and Ni, Bin and Laganenka, Leanid and Sourjik, Victor},
  journal={FEMS microbiology reviews},
  volume={45},
  number={6},
  pages={fuab038},
  year={2021},
  publisher={Oxford University Press}
}

@article{wang2014trophic,
  title={Trophic interactions induce spatial self-organization of microbial consortia on rough surfaces},
  author={Wang, Gang and Or, Dani},
  journal={Scientific Reports},
  volume={4},
  number={1},
  pages={6757},
  year={2014},
  publisher={Nature Publishing Group UK London}
}

@article{caprini2022dynamics,
  title={Dynamics of active particles with space-dependent swim velocity},
  author={Caprini, Lorenzo and Marconi, Umberto Marini Bettolo and Wittmann, Ren{\'e} and L{\"o}wen, Hartmut},
  journal={Soft Matter},
  volume={18},
  number={7},
  pages={1412--1422},
  year={2022},
  publisher={Royal Society of Chemistry}
}

@article{tan2024quasi,
  title={Quasi-two-dimensional dispersions of Brownian particles with competitive interactions: phase behavior and structural properties},
  author={Tan, Zihan and Calandrini, Vania and Dhont, Jan KG and N{\"a}gele, Gerhard},
  journal={Soft matter},
  volume={20},
  number={48},
  pages={9528--9546},
  year={2024},
  publisher={The Royal Society of Chemistry}
}

@article{godfrin2014generalized,
  title={Generalized phase behavior of cluster formation in colloidal dispersions with competing interactions},
  author={Godfrin, P Douglas and Valadez-P{\'e}rez, N{\'e}stor E and Castaneda-Priego, Ramon and Wagner, Norman J and Liu, Yun},
  journal={Soft matter},
  volume={10},
  number={28},
  pages={5061--5071},
  year={2014},
  publisher={The Royal Society of Chemistry}
}

@article{Teubner1987,
  title={Origin of the scattering peak in microemulsions},
  author={Teubner, M. and Strey, R.},
  journal={The Journal of Chemical Physics},
  volume={87},
  number={5},
  pages={3195--3200},
  year={1987},
  publisher={American Institute of Physics}
}

@article{maggi2021universality,
  title={Universality class of the motility-induced critical point in large scale off-lattice simulations of active particles},
  author={Maggi, Claudio and Paoluzzi, Matteo and Crisanti, Andrea and Zaccarelli, Emanuela and Gnan, Nicoletta},
  journal={Soft Matter},
  volume={17},
  number={14},
  pages={3807--3812},
  year={2021},
  publisher={Royal Society of Chemistry}
}

@article{siebert2018critical,
  title={Critical behavior of active Brownian particles},
  author={Siebert, Jonathan Tammo and Dittrich, Florian and Schmid, Friederike and Binder, Kurt and Speck, Thomas and Virnau, Peter},
  journal={Phys. Rev. E},
  volume={98},
  pages={030601},
  year={2018}
}

@article{Hornreich1975,
  author = {R. M. Hornreich and M. Luban and S. Shtrikman},
  title = {Critical Behavior at the Onset of $\vec{k}$-Space Instability on the $\lambda$ Line},
  journal = {Phys. Rev. Lett.},
  volume = {35},
  pages = {1678--1681},
  year = {1975}
}

@article{stenhammar2013continuum,
  author = {Stenhammar, Joakim and Marenduzzo, Davide and Allen, Rosalind J. and Cates, Michael E.},
  title = {Continuum Theory of Phase Separation Kinetics for Active Brownian Particles},
  journal = {Phys. Rev. Lett.},
  volume = {111},
  pages = {145702},
  year = {2013}
}

@article{ng2009bacterial,
  author = {Ng, Wai-Leung and Bassler, Bonnie L.},
  title = {Bacterial quorum-sensing network architectures},
  journal = {Annual Review of Genetics},
  volume = {43},
  pages = {197--222},
  year = {2009}
}

@article{toledano2009colloidal,
  author = {Fernandez Toledano, J. C. and Sciortino, F. and Zaccarelli, E.},
  title = {Colloidal systems with competing interactions: from an arrested repulsive cluster phase to a gel},
  journal = {Soft Matter},
  volume = {5},
  pages = {2390--2398},
  year = {2009}
}

@article{stradner2004equilibrium,
  author = {Stradner, A. and Sedgwick, H. and Cardinaux, F. and Poon, W. C. K. and Egelhaaf, S. U. and Schurtenberger, P.},
  title = {Equilibrium cluster formation in concentrated protein solutions and colloids},
  journal = {Nature},
  volume = {432},
  pages = {492--495},
  year = {2004}
}

@article{zhuang2016recent,
  author = {Zhuang, Yuan and Charbonneau, Patrick},
  title = {Recent Advances in the Theory and Simulation of Model Colloidal Microphase Formers},
  journal = {J. Phys. Chem. B},
  volume = {120},
  pages = {7775--7782},
  year = {2016}
}

@article{zhuang2016equilibrium,
  author = {Zhuang, Yuan and Charbonneau, Patrick},
  title = {Equilibrium Phase Behavior of a Continuous-Space Microphase Former},
  journal = {Phys. Rev. Lett.},
  volume = {116},
  pages = {098301},
  year = {2016}
}

@article{brazovskii1975phase,
  author = {Brazovskii, S. A.},
  title = {Phase transition of an isotropic system to a nonuniform state},
  journal = {Sov. Phys. JETP},
  volume = {41},
  pages = {85--89},
  year = {1975}
}

@article{imperio2006microphase,
  author = {Imperio, A. and Reatto, L.},
  title = {Microphase formation in two-dimensional systems with competing interactions},
  journal = {J. Chem. Phys.},
  volume = {124},
  pages = {164712},
  year = {2006}
}

@article{archer2008twodimensional,
  author = {Archer, Andrew J.},
  title = {Two-dimensional fluid with competing interactions exhibiting microphase separation: Theory for bulk and interfacial properties},
  journal = {Physical Review E},
  volume = {78},
  pages = {031402},
  year = {2008}
}

@article{bates1990block,
  author = {Bates, F. S. and Fredrickson, G. H.},
  title = {Block copolymer thermodynamics: theory and experiment},
  journal = {Annual Review of Physical Chemistry},
  volume = {41},
  pages = {525--557},
  year = {1990}
}

@article{munao2022competition,
  title={Competition between clustering and phase separation in binary mixtures containing SALR particles},
  author={Munao, Gianmarco and Costa, Dino and Malescio, Gianpietro and Bomont, Jean-Marc and Prestipino, Santi},
  journal={Soft Matter},
  volume={18},
  number={34},
  pages={6453--6464},
  year={2022},
  publisher={Royal Society of Chemistry}
}

@article{caprini2018cluster,
  title={Cluster crystals with combined soft-and hard-core repulsive interactions},
  author={Caprini, Lorenzo and Hern{\'a}ndez-Garc{\'\i}a, Emilio and L{\'o}pez, Crist{\'o}bal},
  journal={Physical Review E},
  volume={98},
  number={5},
  pages={052607},
  year={2018},
  publisher={APS}
}

@article{barkan2014controlled,
  title={Controlled self-assembly of periodic and aperiodic cluster crystals},
  author={Barkan, Kobi and Engel, Michael and Lifshitz, Ron},
  journal={Physical review letters},
  volume={113},
  number={9},
  pages={098304},
  year={2014},
  publisher={APS}
}

@article{maggi2022critical,
  title={Critical active dynamics is captured by a colored-noise driven field theory},
  author={Maggi, Claudio and Gnan, Nicoletta and Paoluzzi, Matteo and Zaccarelli, Emanuela and Crisanti, Andrea},
  journal={Communications Physics},
  volume={5},
  number={1},
  pages={55},
  year={2022},
  publisher={Nature Publishing Group UK London}
}

@article{Catesmicrophase,
  doi = {10.1088/1361-6633/add278},
  url = {https://doi.org/10.1088/1361-6633/add278},
  year = {2025},
  month = {may},
  publisher = {IOP Publishing},
  volume = {88},
  number = {5},
  pages = {056601},
  author = {Cates, M. E. and Nardini, C.},
  title = {Active phase separation: new phenomenology from non-equilibrium physics},
  journal = {Reports on Progress in Physics}
}

@article{Joannymicrophase,
  title = {Hydrodynamic theory of chemically active emulsions},
  author = {Ilker, Efe and Laxhuber, Kathrin S. and Joanny, Jean-Fran{\c{c}}ois
            and J{\"u}licher, Frank},
  year = {2026},
  eprint = {2604.17539},
  archivePrefix = {arXiv},
  primaryClass = {cond-mat.soft},
  journal = {arXiv}
}

@article{zhou2024clustering,
  title={Clustering of quorum sensing colloidal particles},
  author={Zhou, Yuxin and Li, Yunyun and Marchesoni, Fabio},
  journal={National Science Open},
  volume={3},
  number={4},
  pages={20230081},
  year={2024},
  publisher={China Science Publishing \& Media Ltd. and EDP Sciences}
}

\clearpage
\onecolumngrid

\setcounter{equation}{0}
\setcounter{figure}{0}
\setcounter{table}{0}
\setcounter{section}{0}
\renewcommand{\theequation}{S\arabic{equation}}
\renewcommand{\thefigure}{S\arabic{figure}}
\renewcommand{\thetable}{S\arabic{table}}
\renewcommand{\theHequation}{S\arabic{equation}}
\renewcommand{\theHfigure}{S\arabic{figure}}
\renewcommand{\theHtable}{S\arabic{table}}
\renewcommand{\theHsection}{S\arabic{section}}

\begin{center}
  {\large\bfseries Supplemental Material:\\[2pt]
   Microphase Separation in Quorum-Sensing Active Particles with Competing Sensing Scales}\\[6pt]
  {Michele Antonioli, Nicoletta Gnan, and Claudio Maggi}
\end{center}
\vspace{1em}


\section{Derivation of the effective Hamiltonian}
\label{sec:effective_hamiltonian}

We report here the derivation of the effective
coarse-grained Hamiltonian used in the main text.

We consider a system of $N$ active particles whose dynamics is
\begin{equation}
    \dot{\mathbf r}_i(t)
    =
    s_i(\mathbf r_1,\ldots,\mathbf r_N)\,
    \boldsymbol \eta_i(t),
\end{equation}
with
\begin{equation}
    \tau \dot{\boldsymbol \eta}_i(t)
    =
    - \boldsymbol \eta_i(t)
    +
    \boldsymbol \xi_i(t),
\end{equation}
where $\boldsymbol \xi_i$ is a Gaussian white noise. The function $s_i$
is a quorum-sensing velocity kernel.

For multiplicative active noise, the stationary probability density of a
single particle at fixed configuration of the others is
\begin{equation}
    p(\mathbf r)
    =
    \frac{1}{Z s(\mathbf r)} .
\end{equation}

Introducing the coarse-grained density field
\begin{equation}
    \rho(\mathbf r)=N p(\mathbf r),
\end{equation}
one obtains the self-consistency equation
\begin{equation}
    \mu
    =
    \log \rho(\mathbf r)
    +
    \log s(\mathbf r),
    \label{eq:self_consistency_SM}
\end{equation}
where $\mu$ plays the role of an effective chemical potential.

The quorum-sensing kernel depends on the perceived density,
\begin{equation}
    s(\mathbf r_i)
    =
    s\!\left(
    \sum_{j\neq i}
    \gamma(|\mathbf r_i-\mathbf r_j|)
    \right).
\end{equation}

In the mean-field approximation, the microscopic density is replaced by its
coarse-grained average,
\begin{equation}
    \sum_j \delta(\mathbf r-\mathbf r_j)
    \longrightarrow
    \rho(\mathbf r),
\end{equation}
so that
\begin{equation}
    s(\mathbf r_i)
    \simeq
    s\!\left(
    \int \dd^d r'\,
    \gamma(|\mathbf r_i-\mathbf r'|)
    \rho(\mathbf r')
    \right).
\end{equation}

We now expand $\rho(\mathbf r')$ around $\mathbf r_i$ up to fourth order.
Since the kernel $\gamma$ is isotropic, all odd terms vanish after
integration. One obtains
\begin{equation}
    s(\mathbf r)
    \simeq
    s\!\left(
    \gamma_0 \rho
    +
    \gamma_2 \nabla^2 \rho
    +
    \gamma_4 \nabla^4 \rho
    \right),
    \label{eq:kernel_gradient_SM}
\end{equation}
where
\begin{equation}
    \gamma_0
    =
    \int \dd^d r\, \gamma(r),
\end{equation}
\begin{equation}
    \gamma_2
    =
    \frac{1}{2d}
    \int \dd^d r\, r^2 \gamma(r),
\end{equation}
and
\begin{equation}
    \gamma_4
    =
    \frac{1}{8d(d+2)}
    \int \dd^d r\, r^4 \gamma(r).
\end{equation}

It is convenient to introduce the rescaled velocity function
\begin{equation}
    v(\rho)
    \equiv
    s(\gamma_0 \rho),
\end{equation}
and the reduced coefficients
\begin{equation}
    g_2=\frac{\gamma_2}{\gamma_0},
    \qquad
    g_4=\frac{\gamma_4}{\gamma_0}.
\end{equation}

Equation~\eqref{eq:self_consistency_SM} then becomes
\begin{equation}
    \mu
    =
    \log \rho
    +
    \log
    v\!\left(
    \rho
    +
    g_2 \nabla^2 \rho
    +
    g_4 \nabla^4 \rho
    \right).
\end{equation}

Expanding the logarithm up to fourth order in gradients gives
\begin{equation}
    \mu
    \simeq
    \log \rho
    +
    \log v(\rho)
    +
    \frac{v'(\rho)}{v(\rho)}
    g_2 \nabla^2 \rho
    +
    \frac{v'(\rho)}{v(\rho)}
    g_4 \nabla^4 \rho .
    \label{eq:mu_gradient_SM}
\end{equation}

We now write the density field as a fluctuation around the homogeneous
state,
\begin{equation}
    \rho(\mathbf r)
    =
    \rho_m+\phi(\mathbf r).
\end{equation}

Expanding Eq.~\eqref{eq:mu_gradient_SM} in powers of $\phi$ gives
\begin{equation}
    \mu
    =
    \mu_0
    +
    a_1 \phi
    +
    a_2 \phi^2
    +
    a_3 \phi^3
    -
    b_2 \nabla^2 \phi
    +
    b_4 \nabla^4 \phi
    +
    \cdots ,
\end{equation}
with
\begin{equation}
    \mu_0
    =
    \log(\rho_m v_m),
\end{equation}
\begin{equation}
    a_1
    =
    \frac{1}{\rho_m}
    +
    \frac{v'_m}{v_m},
\end{equation}
\begin{equation}
    a_2
    =
    \frac{1}{2}
    \left[
    \frac{v''_m}{v_m}
    -
    \left(\frac{v'_m}{v_m}\right)^2
    -
    \frac{1}{\rho_m^2}
    \right],
\end{equation}
\begin{equation}
    a_3
    =
    \frac{1}{6}
    \left[
    \frac{v'''_m}{v_m}
    -
    3\frac{v'_m v''_m}{v_m^2}
    +
    2\left(\frac{v'_m}{v_m}\right)^3
    +
    \frac{2}{\rho_m^3}
    \right],
\end{equation}
and
\begin{equation}
    b_2
    =
    -g_2 \frac{v'_m}{v_m},
    \qquad
    b_4
    =
    g_4 \frac{v'_m}{v_m}.
\end{equation}

Here $v_m=v(\rho_m)$ and derivatives of $v$ are evaluated at $\rho_m$.

The equilibrium-like part of the chemical potential is identified with a
functional derivative,
\begin{equation}
    \mu_{\rm eq}
    =
    \frac{\delta \mathcal H}{\delta \phi}.
\end{equation}

Neglecting non-integrable non-equilibrium terms and retaining only the
symmetric part under $\phi \to -\phi$, the corresponding effective
Hamiltonian is
\begin{equation}
    \mathcal H[\phi]
    =
    \int \dd^d r
    \left[
    \frac{a_1}{2}\phi^2
    +
    \frac{b_2}{2}|\nabla \phi|^2
    +
    \frac{b_4}{2}(\nabla^2\phi)^2
    +
    \frac{a_3}{4}\phi^4
    \right].
    \label{eq:hamiltonian_unscaled_SM}
\end{equation}
After rescaling the Hamiltonian,
Eq.~\eqref{eq:hamiltonian_unscaled_SM}
can be written in the minimal form
\begin{equation}
\mathcal{H}[\phi] = \int \mathrm{d}^{d}r
\left[\frac{c}{2}\phi^{2} + \frac{b}{2}|\nabla\phi|^{2}
+ \frac{1}{2}\left(\nabla^{2}\phi\right)^{2}
+ \frac{\lambda}{4!}\phi^{4}\right],
    \label{eq:hamiltonian_scaled_SM}
\end{equation}
where the phenomenological coefficients are functions of the microscopic
parameters.

In Fourier space, using
\begin{equation}
\phi(\mathbf{r})=\int_{\mathbf{q}}e^{i\mathbf{q}\cdot\mathbf{r}}\,\phi(\mathbf{q}),
\end{equation}
where, here and in the following, $\int_{\mathbf{q}}\equiv\int \mathrm{d}^{d}q/(2\pi)^{d}$,
and $\int_{\mathbf{q}_{1}\ldots\mathbf{q}_{4}}$ denotes the corresponding integration over
the four wavevectors, the quadratic part becomes

\begin{equation}
\mathcal{H}_{2}[\phi]=\frac{1}{2}\int_{\mathbf{q}}
\left(c+bq^{2}+q^{4}\right)\left|\phi(\mathbf{q})\right|^{2}.
\end{equation}

The full Hamiltonian is therefore
\begin{equation}
\begin{split}
\mathcal{H}[\phi] = {}& \frac{1}{2}\int_{\mathbf{q}}
\left(c+bq^{2}+q^{4}\right)\left|\phi(\mathbf{q})\right|^{2}\\
&+\frac{\lambda}{4!}\int_{\mathbf{q}_{1}\ldots\mathbf{q}_{4}}
(2\pi)^{d}\,\delta^{(d)}\!\left(\sum_{i=1}^{4}\mathbf{q}_{i}\right)
\prod_{i=1}^{4}\phi(\mathbf{q}_{i}),
\end{split}
\end{equation}

\noindent where the coefficients follow from Eq.~\eqref{eq:hamiltonian_unscaled_SM} upon rescaling
the Hamiltonian by $b_4$, namely
\begin{equation}
b = \frac{b_2}{b_4} = -\frac{\gamma_2}{\gamma_4}, \qquad
c = \frac{a_1}{b_4}, \qquad
\lambda = \frac{6\,a_3}{b_4}.
\end{equation}


\section{Additional numerical results}
\subsection{State Diagram}
\begin{figure*}[p]
\centering
\safeincludegraphics[width=1.0\textwidth]{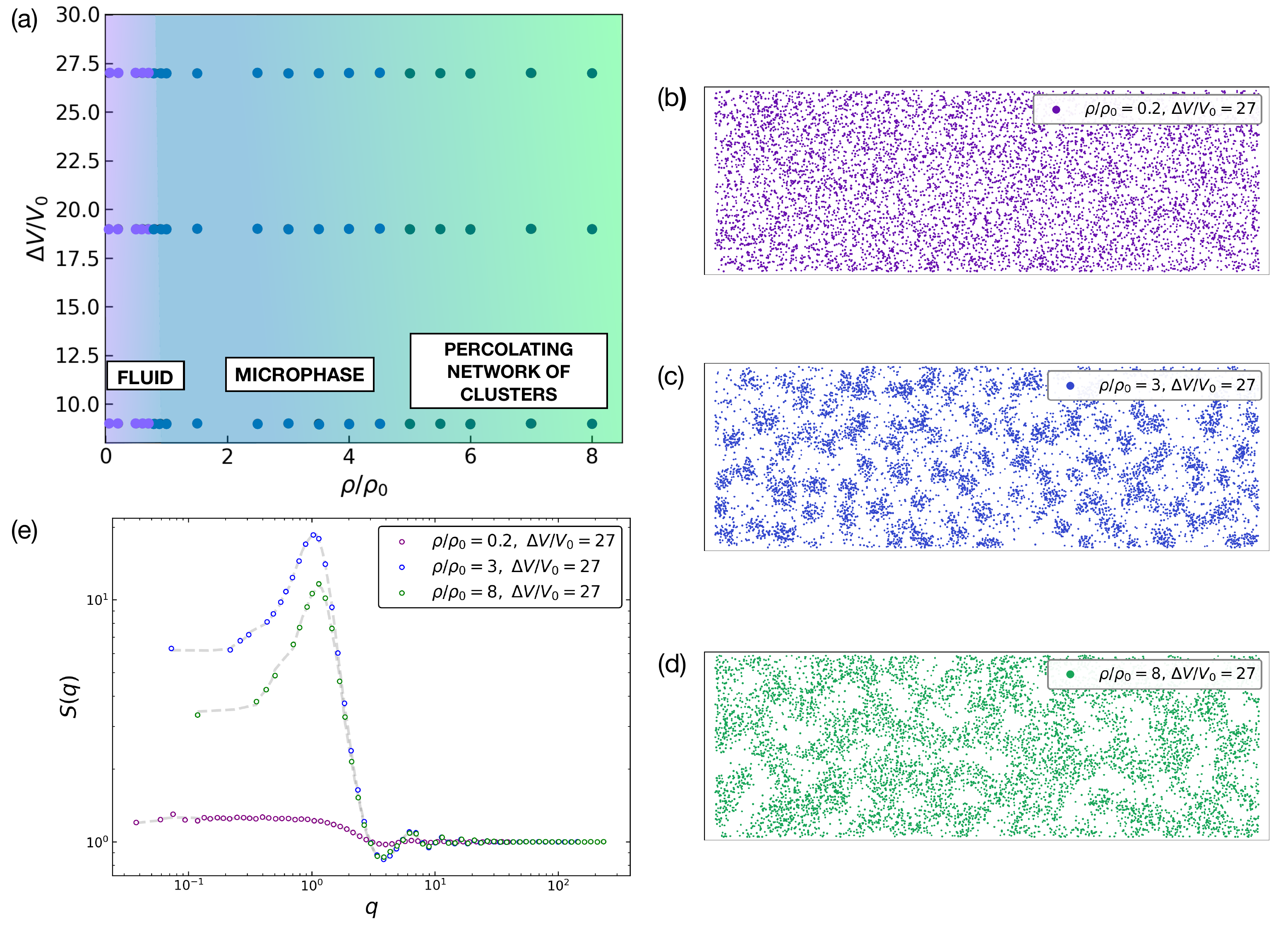}
\caption{\fontsize{11pt}{9pt}\selectfont (a) State diagram obtained by varying the reduced density $\rho/\rho_0$ and the relative speed difference $\Delta V/V_0$ for competing quorum-sensing interactions with $R_1=1.3$ and $R_2=2.6$ ($\gamma_0=1$, $\gamma_2=-0.5$). Three different regimes are identified: a homogeneous fluid at low densities, a microphase-separated state composed of finite-size clusters at intermediate densities, and a percolating network of clusters at high densities. Representative simulation snapshots illustrating the three regimes are shown on the right: (b) fluid state, (c) microphase, and (d) percolating network. (e) Static structure factors of the representative points highlight that, once the finite-wavevector maximum is established, the characteristic modulation length remains essentially unchanged, whereas the particle density controls the connectivity of the resulting microphase.
Note that, although the position of the finite-$q$ peak remains approximately unchanged in the microphase and in the percolating region, its height decreases upon entering the percolating regime. This indicates that the characteristic modulation length is preserved, while the density modulation becomes less pronounced and spatially less coherent as the clusters merge into a percolating network.
\newline\\
\textbf{Alt text:} State diagram, particle configurations, and structure factors across variations in density and motility. Increasing density drives the system from a homogeneous fluid to finite clusters and then to a percolating cluster network, whereas increasing motility mainly enhances the density modulations. The finite-wavevector peak position changes only weakly, indicating that the characteristic modulation length remains nearly constant.
}
 \label{fig:phase_diagram}
\end{figure*}

To assess the robustness of the microphase-separated regime, we
performed a systematic scan over the density $\rho/\rho_0$ and the
relative speed difference $\Delta V/V_0$, fixing the interaction
parameters to $R_1=1.3$, $R_2=2.6$, $\gamma_0=1$, and
$\gamma_2=-0.5$ for a system of $N=15\times10^3$ particles. The
resulting state diagram is shown in Fig.~\ref{fig:phase_diagram}.

At very low densities, the system remains essentially homogeneous over
the whole range of $\Delta V/V_0$ explored (Fig.~\ref{fig:phase_diagram}(b)). Upon increasing the density,
a broad microphase-separated region emerges, characterized by finite
clusters (Fig.~\ref{fig:phase_diagram}(c),(e)). At still higher densities, these clusters progressively merge
into a percolating network while preserving the characteristic
modulation length selected by the competing quorum-sensing
interactions (Fig.~\ref{fig:phase_diagram}(d),(e)).

Interestingly, the overall morphology depends primarily on the particle
density rather than on the speed difference. Increasing
$\Delta V/V_0$ mainly enhances the density contrast between the clusters
and the surrounding dilute phase, whereas the transition from isolated
clusters to a percolating network is largely controlled by the amount of
particles available to populate the modulated structures. This behavior
is consistent with the field-theoretical picture developed in the main
text: the competing interactions determine the characteristic
wavelength through the finite-$q$ minimum of the effective free-energy
functional, while the density controls how particles populate this
preferred modulation, giving rise to different morphologies without
changing the underlying length scale. The broad region occupied by the
microphase states further demonstrates that the finite-wavevector
modulation predicted by the theory is robust over a wide range of
microscopic parameters.

\subsection{Finite-size behavior of the macrophase--microphase transition}
\label{sec:binder}
In order to characterize the transition between macrophase and microphase
separation, we study the finite-size behavior of the Binder cumulant and of the
susceptibility as functions of the control parameter $\gamma_{2}$.
Within the coarse-grained field theory introduced in Sec.~\ref{sec:effective_hamiltonian} and in the main text,
the parameter $\gamma_2$ controls the sign and magnitude of the
square-gradient contribution. Through the mapping derived from the
gradient expansion of the microscopic counting function, $\gamma_2$
directly affects the coefficient $b$ appearing in the effective
Hamiltonian of Eq.~(\ref{eq:hamiltonian_scaled_SM}).
For $b>0$, the quadratic part of the Hamiltonian is minimized at zero wavevector and the
system tends to undergo ordinary macrophase separation, similarly to
standard quorum-sensing MIPS. Conversely, when $b<0$, the minimum of the
quadratic part shifts to a finite wavevector $q^*$,
favoring periodically modulated structures characterized by a finite
emergent length scale. The transition from macrophase to microphase
separation therefore originates from the competition between
long-wavelength ($q=0$) and finite-wavevector ($q=q^\ast$)
instabilities.

This mechanism is reminiscent of the physics of Lifshitz points, where
the dominant instability changes from homogeneous to modulated order as
the square-gradient coefficient changes sign~\cite{Hornreich1975}.
However, the present case differs from the conventional Lifshitz
scenario in an important way. In standard Lifshitz theories, both the
mass term and the square-gradient coefficient are simultaneously tuned
toward a multicritical point. Here, instead, the simulations are performed at fixed density and fixed
$\gamma_{0}$, with $\gamma_{2}$ as the single control parameter.
 The system is therefore not driven through a
conventional critical point, but rather across the boundary between
macrophase separation and microphase formation. 

A closely related framework is provided by Brazovskii
theory~\cite{brazovskii1975phase}. In Brazovskii systems, fluctuations can destroy
the continuous transition predicted at mean-field level and replace it
with a weak fluctuation-induced first-order transition.

The numerical results appear consistent with this scenario.
Indeed, to characterize the transition between macrophase separation and
microphase-separated states, we analyze the behavior of the Binder cumulant defined as 
\begin{equation}
\mathcal{B}
=
\frac{
\langle(\Delta\rho)^2\rangle^2
}{
\langle (\Delta\rho)^4\rangle},
\end{equation}

\noindent where 
\begin{equation}
\Delta\rho
=
\rho_{\rm dense}-\rho_{\rm gas},
\end{equation}
as a function of the microscopic control parameter $\gamma_2$ for
different system sizes.

Here $\rho_{\rm dense}$ and $\rho_{\rm gas}$ are respectively the
average densities measured inside the dense and dilute regions of the
system. To estimate these quantities, we employ the same sub-box
analysis previously used for determining the critical behavior in
standard quorum-sensing active particles~\cite{gnan2022critical,siebert2018critical,maggi2021universality}.
The method exploits the fact that, in the macrophase-separated regime,
the center of mass of the system can be used to identify the dense
domain while excluding the interfacial regions from the analysis.


This procedure works well in the macrophase-separated regime, where the
system displays a well-defined interface separating dense and dilute
regions. Conversely, in the microphase-separated regime the density
field is modulated over finite length scales and no sharp macroscopic
interface is present. In this case, dense and dilute regions coexist
within the same observation boxes, so that the measured local densities
tend to approach a more homogeneous effective value. 
\begin{figure}[h]
\centering
 \safeincludegraphics[width=0.99\linewidth]{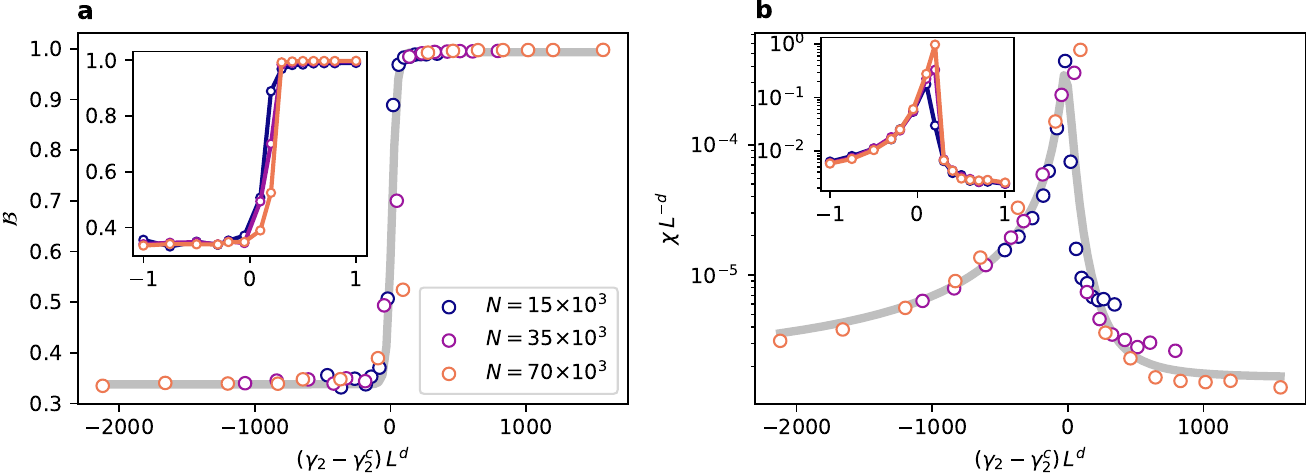}
 \caption{
 \fontsize{11pt}{9pt}\selectfont
Finite-size scaling analysis of the transition between macrophase and
microphase separation. (\textbf{a}) Binder cumulant $\mathcal{B}$ as a function
of the control parameter $\gamma_2$ for different system sizes (inset).
The main panel shows the corresponding finite-size collapse obtained by
plotting $\mathcal{B}$ as a function of
$(\gamma_2-\gamma_2^c)L^d$. 
(\textbf{b}) Susceptibility
$\chi=\langle(N_b-\langle N_b\rangle)^2\rangle/\langle N_b\rangle$ as a
function of $\gamma_2$ (inset). The peak grows with increasing system
size, while the main panel shows the collapse of $\chi/L^d$ as a
function of $(\gamma_2-\gamma_2^c)L^d$, indicating the extensive scaling
$\chi_{\rm max}\sim L^d$ expected for a first-order transition.
\newline\\
\textbf{Alt text:} 
Finite-size scaling of the Binder cumulant and susceptibility
characterizes the macrophase--microphase transition. As system size increases,
the Binder curves sharpen and collapse with volume scaling, without developing a
pronounced minimum, while the susceptibility peak grows extensively, supporting a
weak first-order transition scenario.
}\label{fig:chi_scaling}
\end{figure}

The evolution of $\mathcal{B}$ as a function of $\gamma_2$ for different
system sizes is shown in Fig.~\ref{fig:chi_scaling}(a). As the system
size increases, the Binder cumulant exhibits an increasingly sharp
crossover (inset). The curves
collapse remarkably well when plotted as a function of
$(\gamma_2-\gamma_2^c)L^d$, consistent with the volume scaling expected
for a first-order transition. Nevertheless, unlike strongly first-order
transitions, no pronounced minimum develops upon increasing the system
size. Rather, the Binder cumulant simply becomes progressively steeper,
a behavior consistent with the weakly first-order scenario predicted by
Brazovskii theory.

Further evidence is provided by the susceptibility,
\begin{equation}
\chi=
\frac{\left\langle\left(N_b-\langle N_b\rangle\right)^2\right\rangle}
{\langle N_b\rangle},
\end{equation}
where $N_b$ denotes the instantaneous number of particles contained in a
sub-box of fixed size. The susceptibility for different system sizes is
shown in the inset of Fig.~\ref{fig:chi_scaling}(b). As the transition is
approached, $\chi$ develops a pronounced peak whose height increases
systematically with system size. The susceptibility curves also exhibit
an excellent collapse when rescaled by the system volume and plotted as
a function of $(\gamma_2-\gamma_2^c)L^d$, as shown in the main panel.
This demonstrates that the peak scales extensively,
$\chi_{\mathrm{max}}\sim L^d$, providing further support for the weakly
first-order nature of the transition.

Taken together, the Binder cumulant, the susceptibility, and the field-theoretical results presented in the main text provide a coherent picture. The field theory predicts that the competition between the $q=0$ and finite-$q$ instabilities leads to a fluctuation-driven transition analogous to the Brazovskii scenario, where the transition is expected to be weakly first order. The numerical results are fully consistent with this interpretation: both the Binder cumulant and the susceptibility collapse with volume scaling, while the absence of a pronounced minimum in the Binder cumulant rules out a strongly discontinuous transition, supporting instead the picture of a weak, fluctuation-dominated first-order transition.


\end{document}